\documentclass[floatfix,showpacs,preprintnumbers,amsmath,amssymb]{revtex4}

\usepackage{graphicx}% Include figure files
\usepackage{dcolumn}% Align table columns on decimal point
\usepackage{bm}% bold math

\begin{document}
\title{Kerr-Schild type initial data for black holes with angular momenta}

\author{Claudia Moreno}
\affiliation{Center for Gravitational Physics and Geometry, Penn
State University, University Park, PA 16802}
\email{moreno@gravity.phys.psu.edu}

\author{Dar\'{\i}o
N\'{u}\~{n}ez} \affiliation{Center for Gravitational Physics and
Geometry, Penn State University, University Park, PA 16802}
\affiliation{Instituto de Ciencias Nucleares - UNAM, A.P.
70-543, Ciudad Universitaria, 04510 Mexico, D.F., Mexico}
 \email{nunez@gravity.phys.psu.edu}

\author{Olivier Sarbach}
\affiliation{Department of Physics and Astronomy, Louisiana State University,
202 Nicholson Hall, Baton Rouge, Louisiana 70803--4001}
\email{o.sarbach@alumni.ethz.ch}

\date{\today}

\begin{abstract}
Generalizing previous work
we propose how to superpose spinning black holes in a Kerr-Schild
initial slice. This superposition satisfies several physically
meaningful limits, including the close and the far ones.
Further we consider the close limit of two black holes with opposite
angular momenta and explicitly solve the constraint equations in this case.
Evolving the resulting initial data with a linear code, we
compute the radiated energy as a function of the masses
and the angular momenta of the black holes.
\end{abstract}

  % Numerical simulation; solution of equations, Numerical relativity
                                                % Physics of black holes, Black holes
                                                % 98.62.Mw % Infall, accretion, and accretion disks
\maketitle

\section{Introduction}

The collision of two black holes is one of the most promising
sources for gravitational radiation which the gravitational
observatory LIGO will be trying to detect \cite{LIGO}.

The detection of gravitational waves is a remarkable project which
present challenges in all of their facets, most of them related to
the fact that the signal generated from the merging of two black
holes will arrive to us as a very tiny ripple in spacetime.
Although this is fortunate in the sense that we would not want to
have great spacetime disturbances passing all around us, the fact
that the signal is so feeble makes essential to the project to be
able to theoretically predict and describe with a great deal of
accuracy the gravitational wave generated during the process.

The problem of two interacting black holes can be split into three 
stages, namely: The Newtonian or far limit
when they are already interacting gravitationally, but so far from
each other that the dynamics can be studied within the Newtonian
description. Then, as they get closer, the spacetime begins to be
modified and no simplifying assumptions can be used anymore. It is
supposed that the gravitational distortions will increase. This
period of the evolution is known among the community as the ``full
3D'' part, meaning that the complete Einstein equations have to be
evolved. Finally, in the last stage the two black holes are
suppose to merge. After merging, the resulting black hole is
supposed to wiggle and giggle, eventually settling down to a
stationary black hole. This last part of the process is known as
the close limit \cite{Pullin1}, and its importance resides in the
fact that it can be described as a single perturbed black hole.
In this case the evolution can be carried out using linearized
perturbation theory. For a distorted Schwarzschild
or Kerr black hole, the corresponding equations can be cast in a 
wave-like equation for unconstrained and gauge invariant scalar 
quantities, and the evolution is much more economic, 
numerically speaking, than the full 3D evolution. 

To solve any system of evolution equations one needs initial data.
In general relativity (and gauge theories in general), the
initial data cannot be given freely but has to satisfy some
constraint equations.
There is a wide literature (see \cite{Cook} and references therein)
on how to solve the constraint equations of general relativity.
Most approaches are based
on the York-Lichnerowicz decomposition \cite{York} in which case the 
constraint equations are recast in a set of four coupled elliptic equations.
These equations simplify considerably if one considers conformally
flat initial data, and several interesting solutions which describe a 
slice with two black holes (see \cite{Cook} for a review) have been 
found in this context.
However, solutions with conformally flat three-metric suffer from 
the fact that they cannot exactly represent Kerr black holes since the 
Kerr metric is not known to possess a conformally flat space part 
in the line element, due to
the dragging (for a perturbative proof of nonexistence, see
\cite{GP}.) Therefore, this data is likely to contain ``junk''
radiation which, at least when the two black holes are very far
from each other or have merged and settled down to a stationary
black hole, might not reflect a realistic astrophysical scenario\footnote{Here,
we assume that the final black hole is a Kerr black hole.}.
More recent proposals relax the requirement for the initial data
to be conformally flat \cite{Bis,HMS,Dain} and result in new 
constructions that are able to naturally incorporate spinning black holes.

In the present work, we construct initial data following an approach 
introduced by Bishop {\it et al.} \cite{Bis} and further elaborated 
in \cite{STP, Khanna} which is not directly based on the York-Lichnerowicz 
decomposition. The advantage of this approach is that one can easily obtain
closed analytic expressions for the solution to the constraint equations in 
some limiting cases, including the close limit approximation.
The approach, which is briefly reviewed in Sec. \ref{Sec-KS}, 
uses a Kerr-Schild type of metric in order to construct the initial data. 
While in \cite{Bis} the superposition of two Schwarzschild black holes 
is discussed, we generalize their proposal to spinning black holes in
Sec. \ref{Sec-Super}. Our superposition has some appealing
features (described in more detail in Sec. \ref{Sec-Super}): i)
Far limit: When the two black holes are infinitely far from each
other, one obtains in the region near each one of them, initial
data describing  a single Kerr black hole. ii) Close limit: When
the black holes ``sit on the top of each other'', one obtains
again a single Kerr black hole.

In Sec. \ref{Sec-KKS} we consider the close limit regime in which
two rotating holes are close to each other with their singularities
enclosed by a common apparent horizon. In this case, the initial
slice can be viewed as a distorted $t=const.$ Kerr slice in
Kerr-Schild coordinates. We explicitly solve the linearized constraint
equations for the case in which the angular momenta of the two holes
are opposite but equal in magnitude.
Using a generalized version \cite{ST} of the Zerilli
and Regge-Wheeler equations, we evolve our initial data
in Sec. \ref{Sec-Num} and compute the radiated energy.
In particular, we discuss the dependence of the energy on the
angular momenta of the holes and compare our results to
other calculations based on Bowen-York data \cite{Kha}.
We conclude with a summary and remarks in Sec. \ref{Sec-Conclusion}
and include technical details in appendices \ref{App-A} and \ref{App-B}.

\section{Kerr-Schild initial data}
\label{Sec-KS}

Let us briefly review Bishop {\it et al.}'s proposal \cite{Bis}
to solve the constraint equations of general relativity:
The assumption is that at the initial slice, the metric and its
time derivative are of Kerr-Schild type,
\begin{equation}
ds^2 = ({\eta_{\mu\nu}} + 2\,V\,{l_\mu}\,{l_\nu})dx^\mu dx^\nu,
\label{eq:ks1}
\end{equation}
where $\eta_{\mu\nu}$ is the flat metric,
${\eta_{\mu\nu}}dx^\mu dx^\nu = {dt^2}-dx^2-dy^2-dz^2$,
$V$ is a function and $l_\mu$ a covector which is null,
normalized such that $l_t = -1$ (as a consequence,
$\delta^{ij} l_i l_j = 1$ and $l_i$ has only two independent
components).
The three metric and extrinsic curvature with respect to
a slice $t = $const. are
\begin{eqnarray}
g_{ij} &=& \delta_{ij} - 2V l_i l_j\, ,
\label{eq:gKS}\\
K_{ij} &=&-\frac{1}{\alpha}(V l_i l_j)_{,t} + \alpha[2V l^c(V l_i l_j)_{,c} -
(V l_i)_{,j} - (V l_j)_{,i}],
\label{eq:KKS}
\end{eqnarray}
where $\alpha = 1/\sqrt{1 - 2V}$ is the lapse and where $i = x,y,z$ refer
to Cartesian coordinates.
The way Bishop {\it et al.} propose to solve the constraints is
to make an ansatz for the (spatial part) of the null vector $l_\mu\,$,
to insert the equations (\ref{eq:gKS}) and (\ref{eq:KKS}) into
the Hamilton and momentum constraint equations and to solve
the resulting set of four coupled differential equations for
the unknowns $V$, $\dot{V} \equiv \partial_t V$ and
$\dot{l}_i \equiv \partial_t l_i$. In \cite{Bis}, a
procedure is given for how to solve the nonlinear constraint
equations for the superposition of two non-spinning black holes.
In the close limit approximation (i.e. in the limit where the
distance between the two black holes is much smaller than the mass of the holes)
the constraint equations can even be solved analytically \cite{STP}.

Below, we first generalize Bishop {\it et al.}'s ansatz for $l_i$
to include the superposition of spinning black holes. Then, we
solve the constraint equations in the close limit approximation.
For simplicity, we will only consider the case where the angular momentum
of the two black holes are opposite but equal in magnitude.
In this case, one obtains a perturbed Schwarzschild black hole
which is easier to study than a perturbed Kerr black hole.

\section{Superposing two Kerr black holes}
\label{Sec-Super}

In Cartesian Kerr-Schild coordinates, the Kerr spacetime can
be written in the form (\ref{eq:ks1}) with
\begin{eqnarray}
&V = -\frac{M\,R^3}{R^4+a^2\,z^2}\, ,&  \label{eq:DefV}\\
&l_\mu = \left(-1,-\frac{R\,x+a\,y}{R^2+a^2},-\frac{R\,y-a\,x}{R^2+a^2},-\frac{z}{R}\right),&
\label{eq:lc}
\end{eqnarray}
where $R$ is a positive function with units of
distance given by
\begin{equation}
R^2 = \frac{1}{2}\left( \|\vec{x}\|^2 - a^2
+ \sqrt{(\|\vec{x}\|^2- a^2)^2+4a^2 z^2}\right).
\label{eq:DefR}
\end{equation}
Here, $M$ and $a$ denote the ADM mass and the rotational
parameter, respectively.
Note that $R$ is well-defined outside the disc $\|\vec{x}\|^2 \leq a^2$, $z=0$,
whose boundary is the ring singularity.

In order to generalize Bishop {\it et al.}'s proposal to spinning black holes,
we first notice that the spatial part of the null covector given in Eq. (\ref{eq:lc})
can be rewritten as the sum of two covectors:
\begin{equation}
l_i = l^{(1)}_i + l^{(2)}_i,
\nonumber
\end{equation}
where
\begin{eqnarray}
\vec{l}^{(1)} &=& -\left(\frac{R\,x}{R^2+a^2},\frac{R\,y}{R^2+a^2},\frac{z}{R}
\right),
\nonumber\\
\vec{l}^{(2)} &=& -a\left(\frac{y}{R^2+a^2},-\frac{x}{R^2+a^2},0 \right).
\nonumber
\end{eqnarray}
Here and in the following, we identify $l_i$ with $l^i = \delta^{ij} l_j$
and denote by $\vec{l}$ the vector $(l^x,l^y,l^z)$.
As in the Schwarzschild case \cite{Bis}, the vector $\vec{l}^{(1)}$ can be
written as the gradient of a potential function,
\begin{equation}
\vec{l}^{(1)} = N\,\vec{\nabla}\Phi,
\nonumber
\end{equation}
where $N$ is a normalization function, and
\begin{equation}
\Phi = \frac{M}{R}\, .
\nonumber
\end{equation}
Now, the vector $\vec{l}^{(2)}$, which vanishes in the non-spinning case,
can be written as the cross product of $\vec{l}^{(1)}$, that is, of
$N\,\vec{\nabla}\Phi$ (with the same normalization function), and a
vector proportional to the angular momentum of the black hole:
\begin{equation}
\vec{l}^{(2)} = -N\,\frac{\vec{a}}{R}\times\vec{\nabla}\Phi,
\nonumber
\end{equation}
where $\vec{a}=(0,0,a)$.
If we allow $\vec{a}$ to have components in arbitrary directions and
replace $a\cdot z$ by $\vec{a}\cdot\vec{x}$ in the expressions
(\ref{eq:DefV}) and (\ref{eq:DefR}), we obtain the Kerr solution
for general angular momentum $\vec{J} = M\vec{a}$.

Thus, the spatial part of the covector $l_\mu$ can be written as
\begin{equation}
\vec{l} = N\left( \vec{\nabla}\Phi + \vec{\nabla}\times \vec{b}
\right),
\label{eq:lif}
\end{equation}
where the potential $\Phi$ is given by
\begin{equation}
\Phi = \frac{M}{R}\, ,
\nonumber
\end{equation}
and
\begin{equation}
\vec{b} = \frac{M\,\vec{a}}{2R^2}\, ,
\nonumber
\end{equation}
where now
\begin{equation}
R^2 = \frac{1}{2}\left(
\|\vec{x}\|^2-\|\vec{a}\|^2 + \sqrt{(\|\vec{x}\|^2-
\|\vec{a}\|^2)^2 + 4(\vec{a}\cdot\vec{x})^2}\right).
\nonumber
\end{equation}
Using the linearity of the $\nabla$ operator, we have the
necessary basis to propose the following ansatz for the spatial
part of the null vector, that is, for the description of a Cauchy
slice with characteristics that allow a description of two black holes.

Let the black holes have masses $M_1, M_2$, and angular momenta
$\vec{J}_1 = M_1\vec{a}_1$, $\vec{J}_2 = M_2\vec{a}_2$, respectively
and their centers be at the points $\vec{x}_1$, for black hole one,
and at $\vec{x}_2$, for black hole two. Having written a single Kerr black hole
in the form (\ref{eq:lif}), it is not difficult to generalize Bishop {\it et al.}'s proposal:
We propose
\begin{eqnarray}
\Phi &=& \frac{M_1}{R_1} + \frac{M_2}{R_2},
\label{eq:PhiTwo}\\
\vec{b} &=& \frac{1}{2}\left( \frac{M_1\,\vec{\tilde{a}}_1}{R_1^2}+
\frac{M_2\,\vec{\tilde{a}}_2}{R_2^2} \right),
\label{eq:bTwo}
\end{eqnarray}
with
\begin{eqnarray}
R_i^2 &=& \frac{1}{2}\left( \|\vec{x}-\vec{x}_i\|^2 -
\|\vec{\tilde{a}}_i\|^2+ \sqrt{\left(\|\vec{x} - \vec{x}_i\|^2-
\|\vec{\tilde{a}}_i\|^2\right)^2 + 4\left(\vec{\tilde{a}}_i\cdot
\left(\vec{x} - \vec{x}_i\right)\right)^2}\right),
\label{eq:Ri} \\
\vec{\tilde{a}}_i &=& \left(
1-\vartheta\left(d\right)\right)\,\vec{a}_i + \vartheta\left(d\right)\,\vec{a}_f,
\label{eq:ai}
\end{eqnarray}
where $d=\|\vec{x}_1 - \vec{x}_2\|$ is the ``distance'' between
the black hole's centers, $i=1,2$, and $\vartheta\left(d\right)$ is
an interaction function between the angular momenta of each black
hole. Its specific form can be left undetermined for the moment,
as for the present analysis it is enough that it is a smooth
function which behaves in the following way:
\begin{eqnarray}
\lim_{d\rightarrow 0}\vartheta\left(d\right)\rightarrow &1,&
\nonumber \\
\lim_{d\rightarrow\infty}\vartheta\left(d\right)\rightarrow &0.&
\nonumber
\end{eqnarray}
There are several well known functions with this type of behavior.
The specific form of $\vartheta(d)$ might be related with the
spin-spin interactions of particle physics, and
will be analyzed in future works.

The final rotational parameter $\vec{a}_f$ is given by
\begin{equation}
\vec{a}_f = \frac{M_1\,\vec{a}_1 + M_2\,\vec{a}_2}{M_1 + M_2}\, ,
\label{eq:af}
\end{equation}
which corresponds to the vector addition of the angular momenta,
$\vec{J}_f = \vec{J}_1 + \vec{J}_2$. We will see shortly that Eq.
(\ref{eq:af}) is a necessary condition for obtaining the correct
close limit.

Finally, we comment on the fact that our ansatz has potential
problems at points where
$\vec{k} \equiv \vec{\nabla}\Phi + \vec{\nabla}\times \vec{b}$
vanishes (such that $\vec{l}$ cannot be normalized at these points).
For the superposition of two Schwarzschild black holes, there is
exactly one such point.
For two Kerr black holes with $M_1 = M_2$ and $\vec{a}_1 = -\vec{a}_2$,
we found that $\vec{k}$ is singular at
$\vec{x} = \frac{\vec{x}_1}{\|\vec{x}_1\|} \times \vec{a_1}$,
where the two holes are centered at $\vec{x}_1$, $\vec{x}_2 = -\vec{x}_1$.
In the situation studied in the next section we will see that this
singularity does not cause any problems since there, we can excise a
region which contains the ring singularities of the two holes and
the above singularity.

Besides its mathematical simple form, the above ansatz has
the following useful limits:
\begin{enumerate}
\item{Far limit:  Let us consider that
the center of the black hole 1 is at the origin, and that the
center of the other is far from the origin, that is,
$\|\vec{x}_2\|$ goes to infinity. Since then
$\vartheta\left(d\right)=0$, we obtain from Eq.(\ref{eq:ai}) that
$\vec{\tilde{a}}_i = \vec{a}_i$. So from equation
Eq.(\ref{eq:Ri}), we see that each $R_i^2$ depends only on the
parameters of the respective black hole. Furthermore, in the
vicinity of black hole 1, $R_2$ is very large compared to $R_1$
so the second terms in Eqs.(\ref{eq:PhiTwo},\ref{eq:bTwo}) can be dropped
and one obtains the $l_i$ for one black hole with mass $M_1$ and angular
momentum $\vec{J}_1 = M_1\vec{a}_1$}.
\item{Close limit: Suppose that $\vec{x}_1 = \vec{x}_2\,$, i.e.
the limiting case when the distance $d$ between the two holes goes
to zero. In this case, $\vartheta\left(d\right)=1$, and
$\vec{\tilde{a}}_1=\vec{\tilde{a}}_2=\vec{\tilde{a}}_f$, so $R_1 =
R_2$. Taking into account the expression for the final angular
momentum, Eq.(\ref{eq:af}), the potential $\Phi$, and the vector
$\vec{b}$, Eqs.(\ref{eq:PhiTwo},\ref{eq:bTwo}), we see that one
obtains one black hole with center at $\vec{x}_1$, mass $M=M_1 +
M_2$, and angular momentum $M\,\vec{a}=M_1\,\vec{a}_1 +
M_2\,\vec{a}_2$. }
\item{One hole limit: If the mass of one black hole, say $M_2$, vanishes,
it can be seen from Eq.(\ref{eq:af}) that $\vec{a}_f = \vec{a}_1$
thus, from Eq.(\ref{eq:ai}), $\vec{\tilde{a}}_i=\vec{a}_i$. Thus
the function $R_1$ only depends on $\vec{a}_1$ and the potential
$\Phi$, and the vector $\vec{b}$,
Eqs.(\ref{eq:PhiTwo},\ref{eq:bTwo}), reduce to the expressions
corresponding to black hole 1. }
\item{Schwarzschild limit: Finally, our ansatz reduces to the one
given by Bishop {\it et al.} in the non-spinning case
$\vec{a}_1 = \vec{a}_2 = \vec{0}$. }
\end{enumerate}

\begin{figure}[htb]
\centerline{
\includegraphics[width=8cm]{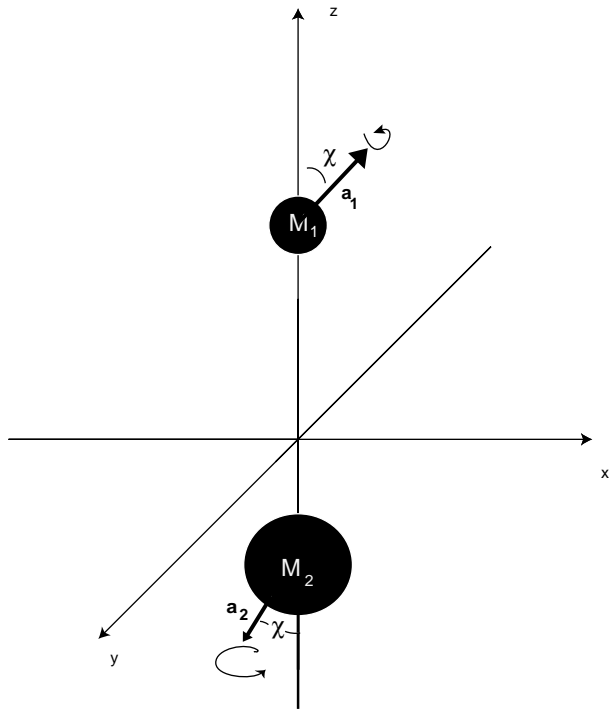}}
\caption{Initial data representing two rotating black holes,
such that the merger is a non rotating one.}
\label{fig:2bh_apz}
\end{figure}

Thus, we consider that we are presenting an educated ansatz for
initial data with certain attractive properties describing the
null vector on a space-like slice with two rotating black holes.
In order to complete the initial data, we have to solve the
constraint equations to yield the function $V$, and the time
derivatives of $V$ and $l_i$. We discuss this below, where we
explicitly solve the constraint equations in the close limit
approximation, where the distance $d$ between the two holes is
infinitesimal. For simplicity, we will also assume that the
angular momenta of the holes are equal in magnitude but opposite,
obtaining a non-spinning black hole when $d = 0$ (see figure 
\ref{fig:2bh_apz}).

\section{Model for ``Kerr+Kerr=Schwarzschild''}
\label{Sec-KKS}

We expand $\Phi$ and $\vec{b}$ in Eqs. (\ref{eq:PhiTwo},\ref{eq:bTwo}),
in terms of the dimensionless ``distance'' between the black holes,
\begin{equation}
\epsilon = \frac{d}{M_1 + M_2}\, ,
\nonumber
\end{equation}
which we consider to be small compared to unity. In the following,
we choose the Cartesian coordinates such that the
two holes are located at $\vec{x}_1 = (0,0,M_2\epsilon)$ and 
$\vec{x}_2 = (0,0,-M_1\epsilon)$, respectively.
This ``center of mass condition'' ($M_1\vec{x}_1 + M_2\vec{x}_2 = \vec{0}$) 
on the coordinate system simplifies the calculations below
since no terms proportional to $\epsilon$ will appear in
the expansions\footnote{The radiated energy, being a coordinate
invariant quantity, does not depend on this coordinate choice.}. 
Similarly, we shall write $\vec{a}_1 = M_2 s\hat{e}$,
and $\vec{a}_2 = -M_1 s\hat{e}$, where $\hat{e}$ is a unit vector,
and $s$ a dimensionless parameter equal to $J_1/M_1 M_2$.
As a consequence of this, $\vec{a}_f$ is equal to zero
(see Eq. (\ref{eq:af})), thus to zeroth order in
$\epsilon$, we have a Schwarzschild black hole.

Using this and the Taylor expansion
\begin{equation}
\vartheta(\epsilon) = 1 + \vartheta^\prime(0)\epsilon +
\frac{1}{2}\vartheta^{\prime\prime}(0)\epsilon^2 + {\cal O}(\epsilon^3),
\nonumber
\end{equation}
for the ``spin--spin interaction'' function,
we eventually arrive at the following expansions for
$\Phi$ and $\vec{b}$
\begin{eqnarray}
\Phi &=& \frac{M}{r} + \frac{M\,M_1\,M_2}{2 r^3}\left[
3\cos^2\theta - 1 + q^2\left(1 - (\hat{e}\cdot\hat{x})^2 \right)
\right]\epsilon^2 + {\cal O}(\epsilon^3),
\label{eq:PhiExp}\\
\vec{b} &=& -\frac{M\,M_1\,M_2}{r^3}\,q\,\cos\theta\,\hat{e}\,\epsilon^2
+ {\cal O}(\epsilon^3),
\label{eq:bExp}
\end{eqnarray}
where $M = M_1 + M_2$ is the total mass, and where we have introduced
$r = \|\vec{x}\|$, $\hat{x} = \vec{x}/r$, $\cos\theta = z/r$.
Here, $q = s\vartheta^\prime(0)$ is proportional to the dimensionless
angular momentum parameter $s$ introduced above. The proportionality
factor depends on the slope of the interpolation function $\vartheta(d)$
at $d=0$. Since $\vartheta(0) = 1$ and $\vartheta(r) \rightarrow 0$ as
$r \rightarrow \infty$ a natural choice for this factor is 
$\vartheta^\prime(0) = -1$ in which case $q=-s$. 
Note that for the superposition of non-extremal
black holes, we have $|s| < 1$.

To summarize, the expansion depends on the parameters: $M_1, M_2$
(which are supposed to model the masses of the black holes), 
$q=-s$ (the dimensionless angular momentum parameter)
and $\hat{e}$ (the orientation of the angular momentum with respect
to the common axis of the holes).

In order to proceed, it is convenient to expand the above
expressions in terms of spherical harmonics. Since the quantities
we consider are real, we choose the real harmonics
$(Y^{\ell 0},R^{\ell m}, I^{\ell m})$, $m = 1,...,\ell$,
defined in Appendix \ref{App-A}.
Parameterizing $\hat{e} = (0,\sin\chi,\cos\chi)$ without loss
of generality and using Eq. (\ref{eq:lif}), the spatial vector
$\vec{l}$ eventually assumes the form
(see Appendix \ref{App-A} for more details)
\begin{eqnarray}
l_r &=& -1 + {\cal O}(\epsilon^3), \nonumber\\
l_A &=& \frac{M_1 M_2}{r}\partial_A\left[
   \gamma_{20}\, Y^{20} + \gamma_{2-1}\, I^{21} + \gamma_{22}\, R^{22}
 + \gamma_{11}\, R^{11} \right]
\epsilon^2 \nonumber\\
 &+& \frac{M_1 M_2}{r}\hat{\varepsilon}^B_{\, A}\partial_B\left[
 \delta_{20}\, Y^{20} + \delta_{2-1}\, I^{21} \right]\epsilon^2
 + {\cal O}(\epsilon^3).
\label{eq:lnf}
\end{eqnarray}
where $A = \theta,\phi$ denote the angular variables
and $\hat{\varepsilon}^B_{\, A}$ denotes the Levi-Civita
tensor with respect to the standard metric on $S^2$.
The constant coefficients $\gamma_{\ell m}$ and $\delta_{l m}$
are given by
\begin{eqnarray}
\gamma_{20} &=& \sqrt{\frac{4\pi}{5}}\left[ 1 + \frac{q^2}{6}
\left( 1 - 3\cos^2\chi \right) \right], \nonumber\\
\gamma_{2-1} &=& -\sqrt{\frac{\pi}{15}}\, q^2\sin(2\chi), \nonumber\\
\gamma_{22} &=& \sqrt{\frac{\pi}{15}}\, q^2\sin^2\chi, \nonumber\\
\gamma_{11} &=& -\sqrt{\frac{4\pi}{3}}\, q\sin\chi, \label{eq:coeff}\\
\delta_{20} &=& \sqrt{\frac{64\pi}{45}}\, q\cos\chi, \nonumber\\
\delta_{2-1} &=& \sqrt{\frac{16\pi}{15}}\, q\sin\chi. \nonumber
\end{eqnarray}

A nice feature of our ansatz is that the only contributions
with $\ell < 2$ are in the even-parity sector with $\ell = 1$,
and this sector is known to contain only gauge modes
(see, for instance, Ref. \cite{ST} for a proof of this statement).
Thus, to first order in $\epsilon^2$,
we have no change in the mass nor in the (total) angular
momentum. It is also interesting to notice that the
even-parity contributions look exactly as in the non-spinning
case (see \cite{Bis, STP}), except that we have to replace
in each sector $M_1 M_2$ by $\gamma_{2 m} M_1 M_2$ (and for
$q=0$ all the $\gamma_{2 m}$'s vanish except for $\gamma_{20}\,$,
in which case we recover the ansatz in the non-spinning case.)

In order to solve the constraint equations, we expand
\begin{eqnarray}
V &=& -\frac{M}{r} + \epsilon^2\left[
v_{20}(r)\, Y^{20} + v_{2-1}(r)\, I^{21} + v_{22}(r)\, R^{22}
\right] + {\cal O}(\epsilon^3),
\nonumber\\
\dot{V} &=& \epsilon^2\left[
\dot{v}_{20}(r)\, Y^{20} + \dot{v}_{2-1}(r)\, I^{21} + \dot{v}_{22}(r)\, R^{22}
\right] + {\cal O}(\epsilon^3),
\label{eq:lnr}\\
\dot{l}_A &=& \epsilon^2\partial_A\left[
   \dot{k}_{20}(r)\, Y^{20} + \dot{k}_{2-1}(r)\, I^{21} + \dot{k}_{22}(r)\, R^{22}
\right] \nonumber\\
 &+& \epsilon^2 \hat{\varepsilon}^B_{\, A}\partial_B\left[
 \dot{n}_{20}(r)\, Y^{20} + \dot{n}_{2-1}(r)\, I^{21} \right] + {\cal O}(\epsilon^3),
\nonumber
\end{eqnarray}
where $v_{2m}(r)$, $\dot{v}_{2m}(r)$ and $\dot{n}_{2m}(r)$ are
unknown functions.
Introducing (\ref{eq:lnf}) and (\ref{eq:lnr}) into the constraint
equations, and keeping only terms of the order $\epsilon^2$,
we obtain a set of linear inhomogeneous differential equations
for these amplitudes, where the inhomogeneity is given by the
coefficients in the expansion (\ref{eq:lnf}).
These equations split into two sets -- one set comprising the
even-parity amplitudes $v_{2m}$, $\dot{v}_{2m}$, $\dot{k}_{2m}$,
and the other set comprising the odd-parity amplitude $\dot{n}_{2m}$.
Amplitudes belonging to different $m$'s also decouple because
the background is spherically symmetric.

As observed above, the coefficients in the expansion of $l_A$ in terms
of spherical harmonics with even parity are proportional to
$\gamma_{2m} M_1 M_2$. Since we have already solved the constraint equations
in the non-spinning case, for which $\gamma_{20} = \sqrt{4\pi/5}$ while
all other $\gamma_{2m}$'s vanish, the solution to the constraint equations
is the same as in the non-spinning case \cite{STP} except that we have to
rescale in each sector $M_1 M_2$ by $\gamma_{2m}$.
The result is
\begin{equation}
v_{2 m}(x) = -\gamma_{2 m}\frac{2\mu}{3}\left( 1 - \frac{2}{x}
 + \frac{3}{x^2} + \frac{3}{x^3} + \frac{6\pi}{x^2} Y_4(i\sqrt{24/x}) \right)
 + \gamma_{2 m}\frac{\mu C_{2m}}{x^2} J_4( i\sqrt{24/x}),
\label{eq:v2m}
\end{equation}
where $x = r/M$, $\mu = M_1 M_2/M^2$ is the dimensionless
reduced mass of the system, and $C_{2m}$ are free constants.
The factor $6\pi$ in front of the Bessel function $Y_4$
ensures that the initial data is asymptotically flat.
The amplitudes $\dot{v}_{2m}$ and $\dot{k}_{2m}$ are
obtained from Eqs. (6) and (7) of Ref. \cite{STP}, where, again,
one has to replace $M_1 M_2$ by $\gamma_{2 m} M_1 M_2$.

In the odd parity sector, the constraints yield
\begin{equation}
\dot{n}_{2m}(x) = \delta_{2m}\frac{\mu \hat{C}_{2m}}{x}\, ,
\nonumber
\end{equation}
where $\hat{C}_{2m}$ are other free constants.

At this point, a question that arises is what meaning can be given to
the free constants $C_{2m}$ and $\hat{C}_{2m}$. As discussed in \cite{Bis, STP},
the constants $C_{2m}$ can be used in order to fix the position of the
apparent horizon (AH) to its background value $x=2$: To leading order in $\epsilon$,
the position of AH is given by the image of the circle
$x = 2$ under the map
\begin{equation}
\vec{x} \mapsto \vec{x} - \epsilon^2\, D(\vec{x})\,\frac{\vec{x}}{x}\; ,
\nonumber
\end{equation}
where the deviation function $D(\vec{x})$ is given by
\begin{equation}
D(\vec{x}) = -\frac{3}{7}
\sum\limits_m \left( \frac{4}{3}\, v_{2m}(x=2) - \gamma_{2m}\mu \right)
Y^{2m}(\vartheta,\varphi).
\nonumber
\end{equation}
The induced metric on the AH is
\begin{equation}
g_{AH} = 4M^2\left( 1 + \epsilon^2\, D(x=2) \right)
\left( d\vartheta^2 + \sin^2\vartheta\, d\varphi^2 \right).
\nonumber
\end{equation}
Since the corresponding scalar curvature is
\begin{equation}
R_{AH} = \frac{1}{4M^2}\left( 2 + 4\epsilon^2 D(x=2) \right),
\nonumber
\end{equation}
we see that the constants $C_{2m}$ describe the deformation of
the AH. There is no physical preferred choice, but for definiteness,
we will choose $C_{2m}$ such that $D(x=2)$ vanishes,
i.e. such that the AH is not deformed.

A geometrical interpretation of the constants $\hat{C}_{2m}$ can be
given as follows: In the presence of an axial Killing field
$\partial_\varphi\,$ ($\varphi\in (0,2\pi)$), we can define the
quantity
\begin{equation}
J(r) = \frac{1}{16\pi} \int_{S_r} \ast\left( d g_{\varphi\mu}
\wedge dx^\mu \right),
\label{Eq:Komar}
\end{equation}
where $S_r$ denotes a spatial sphere of radius $r$ and $\ast$ is the Hodge dual.
For $r\rightarrow\infty$ this gives the Komar angular momentum. For an
axialsymmetric spacetime with Killing horizon at $r = r_H$,
$J(r_H)$ gives the angular momentum of the horizon. For a
Schwarzschild black hole, $J(r=2M) = 0$. Although the Komar
formula does not make sense if $\partial_\varphi$ is not an axial
Killing field, we can linearize Eq. (\ref{Eq:Komar}) around a
Schwarzschild black hole, evaluate the resulting expression at the
AH, and interpret the resulting quantity as the ``angular momentum
of the AH''\footnote{A more precise way to calculate the angular momentum 
of the AH would be to use the corresponding formulas for an isolated horizon
\cite{IH} or the newly introduced notion of dynamical horizon \cite{DH}.
After linearizing the expressions in \cite{IH}, we have obtain the same
result as in Eq. (\ref{Eq:LinKomar}).}. For $\ell=2$, the result is
\begin{equation}
J(r=2M) = \epsilon^2\frac{3M}{\pi}\sum\limits_m \phi_{2m}(x=2)
\int\cos\vartheta\, Y^{2m} d\Omega\, ,
\label{Eq:LinKomar}
\end{equation}
where $\phi_{2m}$ are the gauge invariant amplitudes given in the
next section. However, since $\cos\vartheta$ is proportional to
$Y^{10}$, all the spherical integrals vanish by virtue of the
orthogonality of the spherical harmonics. Nevertheless, it seems
reasonable to associate the gauge-invariant quantities
$\phi_{2m}(x = 2)$ with a measure for the differential rotation
of the AH. Since these quantities depend on $\hat{C}_{2m}$ (see below),
the constants $\hat{C}_{2m}$ can be used in order to control the amount
of this differential rotation.
Again, there is no physically preferred choice for the amount of
differential rotation. For definiteness, we will choose $\hat{C}_{2m}$
such that $\phi(x = 2)$ vanishes.

\section{Numerical Evolution}
\label{Sec-Num}

Here, we evolve the initial data obtained in the previous section
using the linearized Einstein equations. We compute the radiated
energy at infinity and compare with other results. Since the
background metric is Schwarzschild in Kerr-Schild coordinates, we
can use the generalized versions of the equations of Zerilli and
Regge-Wheeler (RW) derived in Ref. \cite{ST} in order to perform the evolution.
Expressing these equations in Kerr-Schild coordinates, we have,
for the case of interest $\ell = 2$,
\begin{eqnarray}
&& \ddot{\psi} - \frac{x-2}{x+2}\psi ^{''} - \frac{2}{x(x+2)}
(\psi' - \dot{\psi }) - \frac{4}{x+2}\dot{\psi '} +
\frac{6(3+6x+4x^2+4x^3)}{x^2(x+2)(3+2x)^2}\psi = 0,
\label{eq:zerilli}\\
&& \ddot{\phi} - \frac{x-2}{x+2}\phi ^{''} - \frac{2}{x(x+2)}
(\phi' - \dot{\phi }) - \frac{4}{x+2}\dot{\phi '} +
\frac{6(x-1)}{x^2(x+2)}\phi = 0,
\label{eq:rw}
\end{eqnarray}
where $\psi$ and $\phi$ denote the Zerilli and the RW amplitude, respectively,
and where a dot and a prime denote derivatives with respect to the
dimensionless coordinates $t/M$ and $x=r/M$, respectively.

In order to evolve these equations, we have to give initial data to
$\psi$, $\dot{\psi}$ and $\phi$, $\dot{\phi}$. We can construct those
from the linearized three metric and extrinsic curvature using the
formulas given in Sec. III of Ref. \cite{ST}.
The three metric and extrinsic curvature
can be obtained from the expressions (\ref{eq:gKS}, \ref{eq:KKS}).
The final result is
\begin{eqnarray}
\psi_{2 m} &=& M\left[ \frac{x^3 v_{2 m} - 6\gamma_{2 m}\mu }{3x(2x + 3)} \right] \, ,
\nonumber\\
\dot{\psi}_{2 m} &=& -M\left[ \frac{ 2x^3(2x - 1) v_{2 m} +
x^4(x - 2) v_{2 m}^\prime + 6\gamma_{2 m}\mu(x-1) }{6x^2(2x+3)} \right]\, .
\nonumber
\end{eqnarray}
where $v_{2m}$ is given in Eq. (\ref{eq:v2m}), and
\begin{eqnarray}
\phi_{2 m} &=& -2M\mu\delta_{2m}\left( \frac{1}{x^2} + \frac{\hat{C}_{2m}}{4x} \right),
\nonumber\\
\dot{\phi}_{2 m} &=& \frac{2M\mu\delta_{2m}}{x^3}\, .
\nonumber
\end{eqnarray}
Here, we choose the constants $C_{2m}$ and $\hat{C}_{2m}$ to be
\begin{equation}
C_{2m} = 2.7, \qquad
\hat{C}_{2m} = -2,
\label{eq:CC}
\end{equation}
which insures that the AH is not deformed and shows no differential rotation,
as discussed in the previous section.

An expression for the radiated energy in terms of gauge-invariant quantities
is given in \cite{ST}. In our case ($\ell = 2$), this energy takes the form
\begin{equation}
\frac{dE}{du} = \frac{3}{2\pi M^2}\sum\limits_m\left( \dot{\psi}_{2 m}^2 + \dot{\phi}_{2 m}^2 \right),
\nonumber
\end{equation}
where $\dot{\psi}$ and $\dot{\phi}$ are evaluated in the radiative zone.
(Note that the radiated angular momentum is zero in our case since both
the initial and the final stage have zero total angular momentum.)
Given the linearity of the equations and the form of the initial data, the
dependence of the time derivatives of the Zerilli and RW functions are
\begin{eqnarray}
\dot{\psi}_{2 m}(t,r) &=& \epsilon^2\mu M\,\gamma_{2 m}\,\dot{\psi}_0(t,r),
\nonumber\\
\dot{\phi}_{2 m}(t,r) &=& \epsilon^2\mu M\,\delta_{2 m}\,\dot{\phi}_0(t,r),
\nonumber
\end{eqnarray}
where the functions $\psi_0$ and $\phi_0$ are dimensionless.
Accordingly, the total radiated energy is
\begin{equation}
E(q,\chi) = M\epsilon^4\mu^2\left[
  c_{even}\,E_{even}
+ c_{odd}\,E_{odd} \right],
\label{eq:Erad}
\end{equation}
where
\begin{eqnarray}
& E_{even} = \int\limits_0^\infty \dot{\psi}_0^2\, \frac{dt}{M}\, ,
\qquad  c_{even} = \frac{3}{2\pi}\left( \gamma_{20}^2 +
\gamma_{2-1}^2 + \gamma_{22}^2 \right),& \nonumber\\
& E_{odd} = \int\limits_0^\infty \dot{\phi}_0^2\, \frac{dt}{M}\, , \qquad
c_{odd} = \frac{3}{2\pi}\left( \delta_{20}^2 + \delta_{2-1}^2
\right).& \nonumber
\end{eqnarray}
Using the expressions (\ref{eq:coeff}), we obtain, after some
simplifications,
\begin{eqnarray}
c_{even} &=& \frac{6}{5}\left[ \left( 1 - \frac{q^2}{3} \right)^2 + q^2\sin^2\chi \right],
\nonumber\\
c_{odd} &=& \frac{8q^2}{15}\left[ 4 - \sin^2\chi \right]. \nonumber
\end{eqnarray}

Since $E_{even}$ and $E_{odd}$ do not depend on the angle $\chi$,
we can easily derive the expressions for the extrema of
the radiated energy, Eq.(\ref{eq:Erad}) with respect to the angle
$\chi$:
\begin{eqnarray}
\frac{\partial E}{\partial\chi} &=& M\epsilon^4\mu^2\,E_{even} \,\frac{6q^2}{5}
\left(1-\frac{4}{9}\Upsilon\right)\sin(2\chi),
\nonumber\\
\frac{\partial^2 E}{\partial\chi^2} &=& M\epsilon^4\mu^2\,E_{even}
\,\frac{12q^2}{5}\left(1-\frac{4}{9}\Upsilon\right)\cos(2\chi),
\nonumber
\end{eqnarray}
where $\Upsilon=\frac{E_{odd}}{E_{even}}$. Thus, we see that
the extrema are at $\chi=0,\frac{\pi}{2}$, and provided that
$E_{even} < \frac{4}{9}E_{odd}$, the maximum of the energy
radiated will be when the angular momentum of the black holes are
parallel to the approaching axis, and the minimum will be when the
angular momentum is perpendicular to it.
Fixing $C_{2m}$ and $\hat{C}_{2m}$ according to Eq. (\ref{eq:CC}),
a numerical evolution using a first order reformulation of
the equations (\ref{eq:zerilli},\ref{eq:rw}) described in
Appendix \ref{App-B} yields (to $1.5\%$ accuracy)
\begin{equation}
E_{even} = 1.74 \times 10^{-4}, \qquad
\frac{4}{9}\, E_{odd} = 1.78 \times 10^{-3},
\nonumber
\end{equation}
so the minimum of the radiated energy is at $\chi = \pi/2$.
The fact that $4E_{odd}/9$ is much larger than $E_{even}$ suggests
that this will also be the case for different values of $C_{2m}$
and $\hat{C}_{2m}$, as long as these do not differ much from the
values chosen here.

\begin{figure}[htb]
\centerline{
\includegraphics[width=6cm]{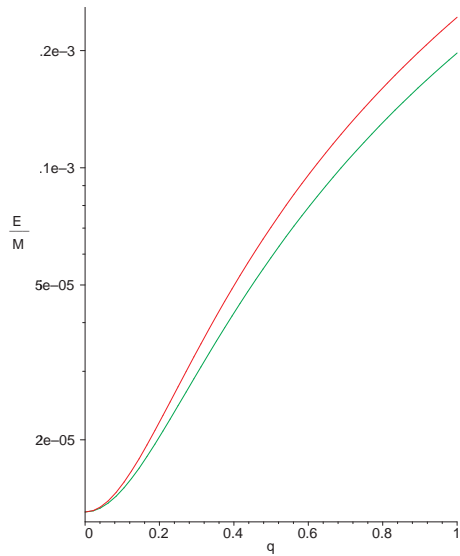}}
\caption{The radiated energy as a function of the
angular momenta for two cases: The upper curve corresponds to the
case when the angular momentum of the black holes are parallel to
the approaching axis ($\chi=0$), and the lower one when it is
perpendicular to it ($\chi=\pi/2$). Here, we have set $\epsilon=1$
and $\mu=1/4$.}
\label{fig2}
\end{figure}

\begin{figure}[htb]
\centerline{
\includegraphics[width=6cm]{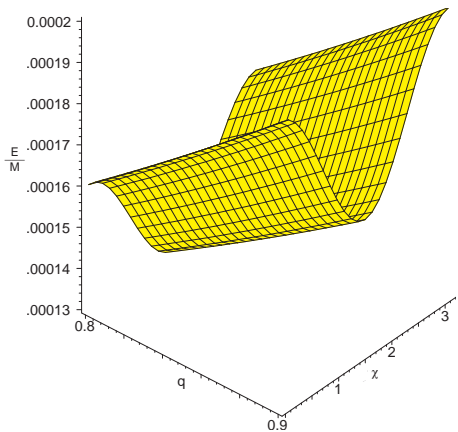}}
\caption{Dependence of the energy radiated on the angle between
the angular momenta of the black holes and the approaching axis,
for values of q close to the extremal case. Here, $\epsilon=1$
and $\mu=1/4$.}
\label{fig3}
\end{figure}

Notice that once $E_{even}$ and $E_{odd}$ have been computed numerically,
the radiated energy can be obtained for {\em any} values of the angular
momentum number $q$ and the angular momentum orientation angle $\chi$ 
by using Eq. (\ref{eq:Erad}).
This function is illustrated in the figures \ref{fig2} and \ref{fig3}. 
The dependence on the angle
$\chi$ is much more noticeable when the angular momentum per square
mass of the black holes is large.
The maximum radiated energy for a fixed total mass and separation
is given by
\begin{equation}
E \leq E(q=1,\chi=0) = M\epsilon^4\times 5.4\times 10^{-4}.
\nonumber
\end{equation}

Our results on the energy radiated agree qualitatively with those
presented recently by Khanna \cite{Kha}. There, the physical situation
sketched in figure \ref{fig:2bh_apz} is modeled using Bowen-York
initial data in the close-slow approximation. It is interesting 
to notice that the quantities we have identified with a measure
for the differential rotation of the AH also vanish in the initial
data used in \cite{Kha} since there, perturbations with odd parity
are trivial.
In particular, Khanna also obtains that there is more
radiated energy for $\chi = 0$ than for $\chi=\pi/2$.
However in \cite{Kha}, the initial data depends
linearly on $q$, so the radiated energy depends only quadratically on $q$
whereas in our initial data, the dependence on $q$ is more complicated.
In particular, we have a contribution in both the even and the odd parity
sectors.

\section{Summary and concluding remarks}
\label{Sec-Conclusion}

We have made an ansatz to give initial data representing the
superposition of two Kerr black holes. The ansatz depends on the
parameters $M_i$ (masses), $\vec{J}_i$ (angular momentum), and an
(unphysical) distance $d$ between the two black holes. The ansatz
is such that when $d$ goes to infinity, one obtains two Kerr
black holes in Kerr-Schild coordinates with masses $M_i$ and
angular momentum $\vec{J}_i$. Furthermore, when
$d=0$, one obtains one Kerr black hole in Kerr-Schild coordinates
with parameters $M = M_1 + M_2$, and $\vec{J} = \vec{J}_1 +
\vec{J}_2$.

The ansatz superposes two black holes with vanishing linear momentum,
and a natural question is whether or not we can generalize our proposal to
boosted black holes. In their paper \cite{Bis}, Bishop {\it et
al.} indicate how to superpose two boosted Schwarzschild black
holes: There, one can replace the potential $\Phi(\vec{x}) =
1/|\vec{x}|$ by a more complicated expression
$\Phi_{\vec{v}}(\vec{x})$ that depends on the speed, $\vec{v}$, of
the boost. A proposal for two boosted non-spinning black holes
which is along the lines of this article would be to take
\begin{equation}
\Phi(\vec{x}) = M_1\Phi_{\vec{\tilde{v}}_1}(\vec{x} - \vec{x}_1) +
M_2\Phi_{\vec{\tilde{v}}_2}(\vec{x} - \vec{x}_2),
\nonumber
\end{equation}
where $\vec{\tilde{v}}_j$ are defined in a similar way as the
$\vec{\tilde{a}}_j$ above. Whether or not this can easily be
generalized to spinning black holes remains an open question.

For the case of two un-boosted Kerr black holes which have
opposite angular momentum but which are equal in magnitude,
we explicitly solved the
constraint equations in the limit in which the black holes
are close to each other. When solving the constraint equations,
integration constants appear. These are associated with a deformation
and a differential rotation of the apparent horizon.
Although we have a geometrical picture for them,
we do not know which values these constants will take
in an astrophysical realistic collapse. It should be stressed
that this is a generic problem of close limit calculations 
since realistic data can only be provided once the full 3D part of 
a black hole collision is solved.
Fixing the integration constant to some specific value,
we evolved the resulting distorted Schwarzschild black hole
using the generalized Zerilli and RW equations in arbitrary coordinates
and computed the radiated energy. Our result might change if
the constants are chosen differently, but nevertheless, when
compared to results with a different choice of initial slice,
we found a qualitatively similar behavior of the radiated
energy as a function of the angular momentum of the black holes.
In particular, we confirmed that there is more energy radiated
when the angular momentum of the black holes are parallel to the
approaching axis.

A similar analysis to the one presented here, but for the far
limit case, is already underway and will be helpful in giving us a
clearer physical understanding of the parameters appearing in
our initial data, in particular in view of post-Newtonian calculations.
Also, the physical meaning of our initial data in the intermediate
region remains to be understood.

Finally, we plan to repeat our results using a version of the
Teukolsky equation in horizon penetrating coordinates \cite{MN},\cite{MNS}
in order to test the latter and in order to proceed with
more confidence to the case where the final black hole is
a rotating one.

\section{Acknowledgments}
We thank Pablo Laguna for the support given during the elaboration
of the present work, specially with the numerical code.
We also thank Badri Krishnan and Manuel Tiglio for many
helpful discussions.
CM and DN acknowledge Mexican Council of Science and Technology, CONACyT,
grants for partial support, DN also acknowledges the DGAPA-UNAM
grant for partial support. OS does so for the Swiss National
Science Foundation.

\appendix

\section{Harmonic decomposition}
\label{App-A}

When expanding real quantities defined on the two-sphere $S^2$,
it is convenient to work with the real spherical harmonics.
In terms of the standard harmonics $Y^{\ell m}$, these
are defined by
\begin{eqnarray}
R^{\ell m} &=& \frac{1}{\sqrt{2}}\left( Y^{\ell m} + Y^{\ell-m} \right) ,
\nonumber\\
I^{\ell m} &=& \frac{1}{\sqrt{2}i}\left( Y^{\ell m} - Y^{\ell-m} \right).
\nonumber
\end{eqnarray}
As the standard harmonics, the real
harmonics defined here provide an orthonormal basis of square
integrable functions on $S^2$.

For $\ell = 1$, we have
\begin{eqnarray}
Y^{10} &=& \sqrt{\frac{3}{4\pi}}\,\cos\theta,
\nonumber\\
R^{11} &=& \sqrt{\frac{3}{4\pi}}\,\sin\theta\cos\phi,
\nonumber\\
I^{11}&=& \sqrt{\frac{3}{4\pi}}\,\sin\theta\sin\phi.
\nonumber
\end{eqnarray}
Note that with respect to this, the radial unit vector $\hat{x}$
can be written as $\hat{x} = \sqrt{4\pi/3}(R^{11},I^{11},Y^{10})$.
For $\ell = 2$, the real harmonics are
\begin{eqnarray}
Y^{20} &=& \sqrt{\frac{5}{4\pi}}\,\frac{1}{2}\left(3\cos^2\theta - 1 \right),
\nonumber\\
R^{21} &=& \sqrt{\frac{15}{4\pi}}\,\sin\theta\cos\theta\cos\phi,
\nonumber\\
I^{21}&=& \sqrt{\frac{15}{4\pi}}\,\sin\theta\cos\theta\sin\phi,
\nonumber\\
R^{22} &=& \sqrt{\frac{15}{16\pi}}\,\sin^2\theta\cos(2\phi),
\nonumber\\
I^{22}&=& \sqrt{\frac{15}{16\pi}}\,\sin^2\theta\sin(2\phi).
\nonumber
\end{eqnarray}

Choosing $\hat{e} = (0,\sin\chi,\cos\chi)$, and expanding the
expressions for $\Phi$ and $\vec{b}$ given in (\ref{eq:PhiExp},\ref{eq:bExp})
in terms of the harmonics above, we obtain
\begin{eqnarray}
\Phi &=& \frac{M}{r} + \frac{M\,M_1\,M_2}{r^3}\,\sqrt{4\pi}\left\{
\frac{q^2}{3} Y^{00}
 + \frac{1}{\sqrt{5}}\left[ 1 + \frac{q^2}{6}(1 - 3\cos^2\chi) \right] Y^{20}
\right. \nonumber\\
&& \left. - \frac{1}{\sqrt{15}}\,\frac{q^2}{2}\left[ 2\sin\chi\cos\chi I^{21}
  - \sin^2\chi R^{22} \right] \right\}\,\epsilon^2 + {\cal O}(\epsilon^3),
\nonumber\\
\vec{b}\cdot d\vec{x} &=& -\frac{M\,M_1\,M_2}{r^3}\,q\,\sqrt{\frac{4\pi}{3}}\left\{
\left[ \frac{\cos\chi}{\sqrt{3}} Y^{00} + \sqrt{\frac{4}{15}}\cos\chi Y^{20}
 + \frac{\sin\chi}{\sqrt{5}} I^{21} \right] dr \right.
\nonumber\\
&& \left. + \frac{r}{2}\,\sin\chi\left[ \frac{1}{\sqrt{5}}\, dI^{21} + \hat{\ast} dR^{11} \right] +
\frac{r}{\sqrt{15}}\,\cos\chi\, dY^{20} \right\}\, \epsilon^2
+ {\cal O}(\epsilon^3).\,
\nonumber
\end{eqnarray}
where $\hat{\ast}$ denotes the Hodge dual on $S^2$.
Using the above expressions and rewriting Eq. (\ref{eq:lif})
as
\begin{equation}
\vec{l}\cdot d\vec{x} = N\left( d\Phi + \ast d(\vec{b}\cdot d\vec{x}) \right),
\nonumber
\end{equation}
where $\ast$ denotes the Hodge dual on the three dimensional Euclidean space
and $N$ is chosen such that $\vec{l}$ is normalized,
it is not difficult to find the expressions given in (\ref{eq:lnf}).

\section{Numerical evolution}
\label{App-B}

Let us consider the following general one dimensional second order
differential equation:
\begin{equation}
\partial_t^{\, 2} u + 2f_1\,\partial_t\partial_x u + f_2\,\partial_x^{\, 2} u
+ f_3\,\partial_t u + f_4\,\partial_x u + f_5\,u + f_6=0,
\label{eq:ap1}
\end{equation}
with $u$ and $f_i$ functions of $(t,x)$.
We want to reformulate this equation as a system which is first order
in time and space. To this purpose, we introduce the following new variable:
\begin{equation}
\pi = (\partial_t + g_+\,\partial_x)\,u,
\nonumber
\end{equation}
with $g_\pm$ to be determined below. We have
\begin{equation}
(\partial_t + g_-\,\partial_x)\,\pi = \left\{
\partial_t^{\, 2} + (g_++g_-)\,\partial_t\partial_x + g_+\,g_-\,\partial_x^{\, 2}
  + \left[ (g_+)_{,t}+\,g_-\,(g_+)_{,x} \right]\partial_x
\right\} u.
\nonumber
\end{equation}
Comparing this last expression with Eq. (\ref{eq:ap1}), we see that if
\begin{equation}
g_\pm = f_1 \pm \sqrt{f_1^2 - f_2}\, ,
\nonumber
\end{equation}
the original system (\ref{eq:ap1}) assumes the form
\begin{equation}
\bf{\partial_t v = A\partial_x v + B\,v + S},
\label{eq:asf}
\end{equation}
with
\begin{equation}
\bf{v} = \left( \begin{array}{c c} u \\ \pi \end{array} \right), \qquad
\bf{A} = \left( \begin{array}{c c} -g_+ & 0 \\ A_0 & -g_- \end{array} \right),\qquad
\bf{B} = \left( \begin{array}{c c} 0 & 1 \\ -f_5 & -f_3 \end{array} \right), \qquad
\bf{S} = \left( \begin{array}{c c} 0 \\ -f_6 \end{array} \right),
\nonumber
\end{equation}
where
\begin{equation}
A_0 = (g_+)_{,t} + g_-\,(g_+)_{,x} + f_3\,g_+ - f_4\, .
\nonumber
\end{equation}

Applying the above to the equations (\ref{eq:zerilli},\ref{eq:rw}),
we find, in both cases,
\begin{equation}
g_+ = \frac{x-2}{x+2}\, ,\qquad
g_- = -1, \qquad
A_0 = 0.
\nonumber
\end{equation}
Since $\bf{A}$ is diagonal with real eigenvalues,
the evolution system (\ref{eq:asf}) is strongly hyperbolic.
One of the advantages of having a strongly hyperbolic system is
that there is a well-defined way how to give boundary conditions
(see, for example, \cite{KL}):
$u$ and $\pi$ are characteristic modes with characteristic speeds
$-g_+$ and $-g_-$, respectively. Inside the horizon ($x < 2$),
both speeds are positive which means that all modes leave the domain
if we put the inner boundary inside the horizon.
Thus, we can excise the singularity and extrapolate $u$ and $\pi$ at
the inner boundary, which we put at $x < 2$.
As long as $x > 2$, $u$ is a mode with positive speed and $\pi$
a mode with negative speed. At the outer boundary, we impose the
outgoing wave condition $\pi = 0$.

We have written a code that numerically evolves (\ref{eq:asf}).
This code is a second order accurate finite differencing one.
Spatial derivatives are discretized using centered differencing
and the time update is performed with an iterated Crank-Nicholson
algorithm using two iterations. We have tested our code by comparing
the resulting waveforms with the ones produced by the code used in
Ref. \cite{STP}.
We have calculated the quantities $E_{even}$ and $E_{odd}$ defined
in Sec. \ref{Sec-Num} by integrating the Zerilli and Regge-Wheeler
function, respectively, over time at a fixed location $r_{obs}$ far 
away from the event horizon of the Schwarzschild background black hole. 
We have checked that the numerical results for $E_{even}$ and $E_{odd}$ 
at fixed $r_{obs}$ converge as numerical resolution is increased and
that these results do not change by more than $1.5\%$ when $r_{obs}$ is increased
(with typical values of $r_{obs}$ lying in the range $101M \leq r_{obs} \leq 151M$.)


\begin{thebibliography}{99}

\bibitem{LIGO}
See the site http://ligo.caltech.edu, for the most recent advances on the LIGO.

\bibitem{Pullin1}
R. Price and J. Pullin,
Phys. Rev. Lett. \textbf{72}, 3297 (1994);
P. Anninos, R. Price, J. Pullin, E. Seidel, and W.-M. Suen,
Phys. Rev. D \textbf{52}, 4462 (1995).

\bibitem{Cook}
G. Cook,
Living Rev.\ Rel.\  \textbf{3}, 5 (2000).

\bibitem{York}
J.W. York Jr.,
J. Math. Phys., \textbf{14}, 456 (1973). 

\bibitem{GP}
A. Garat and R. H. Price,
Phys. Rev. D \textbf{61}, 124011 (2000).

\bibitem{Bis}
N. T. Bishop, R. Isaacson, M. Maharaj, and J. Winicour,
Phys. Rev. D \textbf{57}, 6113 (1998).

\bibitem{HMS}
R.~A.~Matzner, M.~F.~Huq and D.~Shoemaker,
Phys.\ Rev.\ D \textbf{59}, 024015 (1999).

\bibitem{Dain}
S. Dain,
Phys. Rev. Lett. \textbf{87}, 121102 (2001).

\bibitem{STP}
O. Sarbach, M. Tiglio, and J. Pullin,
Phys. Rev. D \textbf{65}, 064026 (2002).

\bibitem{Khanna}
G. Khanna,
Phys.Rev. D \textbf{65}, 124018 (2002). 

\bibitem{ST}
O. Sarbach and M. Tiglio,
Phys. Rev. D \textbf{64}, 084016 (2001).

\bibitem{Kha}
G. Khanna,
Phys. Rev. D \textbf{63}, 124007 (2001).

\bibitem{IH}
A. Ashtekar, C. Beetle, O. Dreyer, S. Fairhurst, B. Krishnan, J. Lewandowski, and J. Wisniewski,
Phys. Rev. Lett. \textbf{85}, 3564 (2000);
A. Ashtekar, C. Beetle, and J. Lewandowski,
Class. Quantum Grav. \textbf{19}, 1195 (2002);
Phys. Rev. D \textbf{64}, 044016 (2001).

\bibitem{DH}
A. Ashtekar and B. Krishnana,
\textit{Dynamical horizons: energy, angular momentum, fluxes and balance laws},
gr-qc/0207080. 

\bibitem{MN}
C. Moreno and D. N\'{u}\~{n}ez,
Int. J. of Mod. Phys. D \textbf{11}, 1331 (2002).

\bibitem{MNS}
C. Moreno, D. N\'{u}\~{n}ez, and O. Sarbach,
in preparation.

\bibitem{KL}
H. Kreiss and J. Lorentz,
\textit{Initial boundary value
problems and the Navier Stokes equation},
Academic Press, San Diego, (1983).


\end{thebibliography}
\end{document}